\documentclass[twocolumn, 10pt]{article}
\usepackage{graphicx,color}  
\usepackage{sidecap} 
\usepackage{dcolumn}   
\usepackage{bm}        
\usepackage{amssymb}   

\usepackage{ulem}
\usepackage{amsmath}
\usepackage{epstopdf}
 
\usepackage{siunitx} 

\newcommand{\nm}{\,\si{\nano\meter}}
\newcommand{\um}{\,\si{\micro\meter}}

\newcommand{\uL}{\,\si{\micro\liter}}

\newcommand{\uM}{\si{\micro\textsc{m}}}
\newcommand{\nM}{\si{\nano\textsc{m}}}

\usepackage[margin=2 cm]{geometry}
\usepackage[textsize=tiny]{todonotes}

\usepackage{authblk}

\title{A simple method to reprogram the binding specificity of DNA-coated colloids that crystallize}

\author[1]{Pepijn G. Moerman}
\author[2]{Huang Fang}
\author[2]{Thomas E. Videb\ae k}
\author[2,*]{W. Benjamin Rogers}
\author[1,$\dagger$]{Rebecca Schulman}
\affil[1]{Department of Chemical and Biomolecular Engineering, Johns Hopkins University, Baltimore, MD 21218, USA}
\affil[2]{Martin A. Fisher School of Physics, Brandeis University, Waltham, MA 02453, USA}
\affil[*]{Correspondence can be addressed to: wrogers@brandeis.edu}
\affil[$\dagger$]{Correspondence can be addressed to: rschulm3@jhu.edu}
\date{}                     
\begin{document}
  \maketitle


\begin{abstract}
\textbf{DNA-coated colloids can crystallize into a multitude of lattices, ranging from face-centered cubic to diamond and thereby contribute to our understanding of crystallization and open avenues to producing structures with useful photonic properties. Despite the broad potential design space of DNA-coated colloids, the design cycle for synthesizing DNA-coated particles is slow: preparing a particle with a new type of DNA sequence takes more than one day and requires custom-made and chemically modified DNA that typically takes the supplier over a month to synthesize. Here, we introduce a method to generate particles with custom sequences from a single feed stock in under an hour at ambient conditions. Our method appends new DNA domains onto the DNA grafted to colloidal particles based on a template that takes the supplier less than a week to produce. The resultant particles crystallize as readily and at the same temperature as those produced via direct chemical synthesis. Moreover, we show that particles coated with a single sequence can be converted into a variety of building blocks with differing specificities by appending different DNA sequences to them. This approach to DNA-coated particle preparation will make it practical to identify optimal and complex particle sequence designs and to expand the use of DNA-coated colloids to a much broader range of investigators and commercial entities.}
\end{abstract}

\maketitle

\section{Introduction}
\begin{figure}
\includegraphics[scale=1]{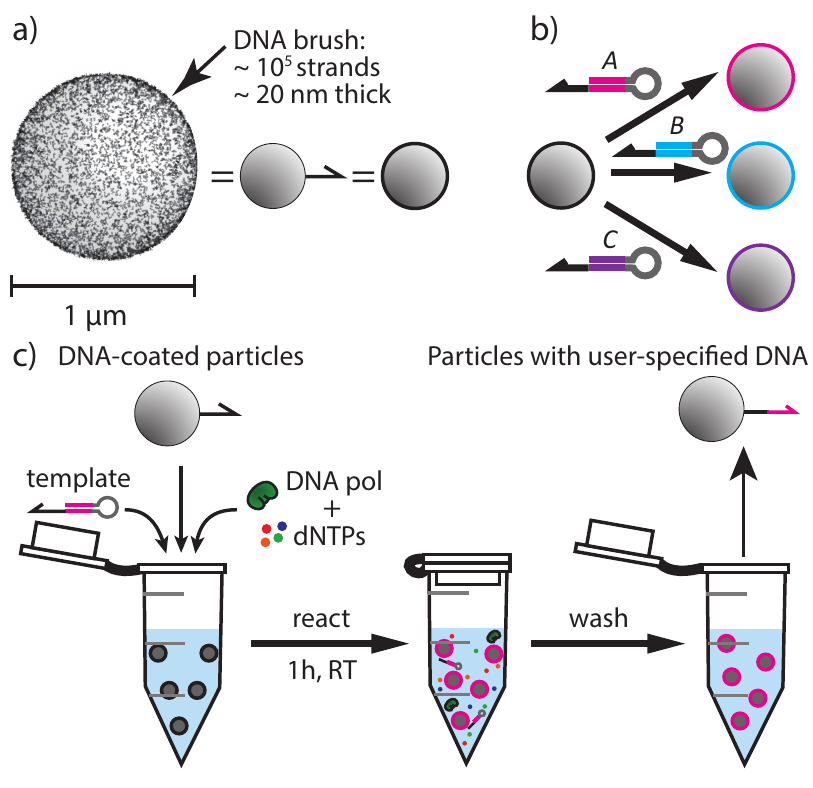}
\caption{\label{Fig0} a) Schematic of DNA-coated colloids. Cartoon of DNA brush on colloid is copied from Ref~\cite{Rogers2016} b) The primer exchange reaction enables the production
of a range of DNA-coated colloids with distinct binding
specificities from a single particle feed stock. c) Overview of the primer exchange reaction (PER) that extends the DNA on DNA-coated particles. DNA polymerase, a desoxynucleotide triphosphate (dNTP) mixture, and a DNA sequence template are mixed in an Eppendorf tube and left at room temperature. Then the particles are separated from the reaction mixture by centrifugation, at which point they are ready to be used in self-assembly experiments. }
\end{figure}

Due to the specificity of DNA hybridization, orthogonal interactions can be prescribed between microscopic objects by coating the objects with orthogonal, complementary pairs of single-stranded DNA\cite{Rogers2016,Mirkin1996,Alivisatos1996}; building blocks with complementary sequences have short-ranged attractive interactions resulting from the hybridization of the DNA on their surfaces\cite{Rogers2011}. This use of DNA is an established strategy for producing building blocks that can assemble into a wide variety of microscopic structures, including stick figures~\cite{Liu2016}, crystal lattices~\cite{Park2008,Auyeung2014,Rogers2015,Wang2017}, flexible bead-chains~\cite{McMullen2018,Verweij2020}, chiral clusters~\cite{BenZion2017}, and even cell aggregates~\cite{Todhunter2015}. Because  DNA-coated microparticles (Fig. 1a) have sizes comparable to the wavelength of visible light, they are particularly promising building blocks for the self-assembly of photonic bandgap materials~\cite{Cersonsky2021,He2020,Ducrot2017,Hensley2022}, with applications in optical wave guides, lasers, and various light-harvesting technologies. DNA-coated microparticles are also useful as model systems for self-assembly, both in~\cite{Oh2019} and out of equilibrium~\cite{Zeravcic2014,Bausch2019}. 

DNA can be grafted onto colloidal particles in various ways, but not all methods produce particles that are compatible with equilibrium assembly of colloidal crystals~\cite{Moon2018}. When biotin-streptavidin chemistry is used to attach single-stranded DNA to particles, the particles tend to hit-and-stick and become kinetically trapped in fractal-like aggregates, even at temperatures at which the DNA-mediated interactions are reversible~\cite{Kim2006, Vitelli2013}. Attaching DNA to particles using strategies based on strain-promoted click chemistry~\cite{Wang2015a,Oh2015,Oh2020} produce DNA-coated colloids that crystallize~\cite{Wang2015b,Fang2020,Hensley2022}. However, these click-chemistry-based methods are time-consuming and require specialized knowledge of synthetic chemistry, which stands in the way of the widespread use of DNA-coated colloids. Moreover, these methods require dibenzocyclooctyne (DBCO)-functionalized DNA, which currently takes roughly a month to be commercially synthesized\cite{IDT}, so that---even if one has the expertise necessary for this synthesis---the time between the initial idea and the experiment is over a month. Last, once the particles are synthesized, their DNA sequences and thus their specificities for other particles are fixed. New particles must therefore be synthesized for each experiment or application that requires a unique DNA sequence.

\begin{figure}
\includegraphics[scale=1]{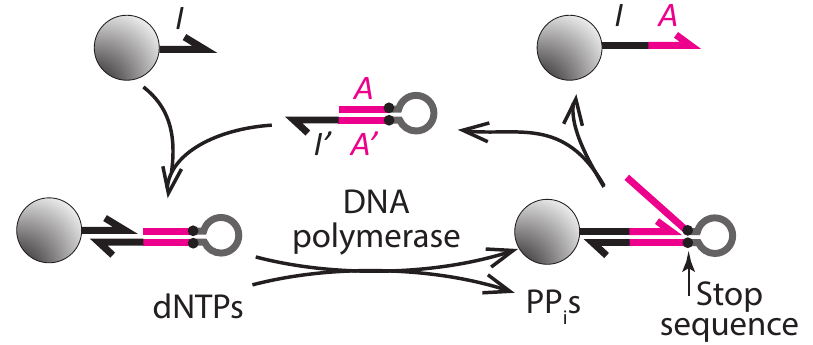}
\caption{\label{Fig1} Schematic of the primer exchange reaction. dNTPs are DNA nucleotides and PP$_i$s are inorganic pyrophosphates. The input sequence, \textit{I}, is 9 nucleotides, and the output domain, \textit{A}, is 11 nucleotides. Sequences are in the Supplementary Information (SI section 1.6). To stop the DNA polymerase from copying more DNA after it has copied the output domain, an artificial stop sequence is incorporated in the template. This stop sequence consists of two nucleotides of which the complement cannot be incorporated because the correponding NTP is not present in the reaction mixture. For example, in a solution that lacks GTP, the incorporation of a G stops the polymerase. More details on designing stop sequences are provided in the SI (section 2).}
\end{figure}

Here we introduce a simple method to synthesize DNA-coated particles with user-prescribed sequences from a single particle feed stock (Fig. 1b). Our method decouples the expensive and time-consuming step of attaching DNA to colloidal particles from the step of tailoring the DNA sequence for its particular purpose, enabling one to convert the sequence on a batch of DNA-coated particles into another sequence for each new experiment, rather than redoing the DNA-coating procedure. Our method uses the Primer Exchange Reaction (PER), introduced by Kishi \textit{et al.} in 2019~\cite{Kishi2019}, to append a user-specified domain to the end of an input DNA sequence coating the particles. We show that this reaction can completely convert the particles' DNA sequence within 1 hour (Fig. 1c) and that particles synthesized using this method---from now on referred to as PER-edited particles---assemble as readily and have the same melting temperature as particles produced directly \textit{via} click chemistry---from now on referred to as reference particles. We also show that a single type of DNA-coated particle can be converted into a variety of particles with different DNA sequences and binding specificities (Fig. 1b), thereby overcoming some of the key technical bottlenecks to the synthesis of DNA-coated colloids and facilitating their widespread use. 
   
\section{Results}
\subsection{Conversion}

\begin{figure}
\includegraphics[scale=1]{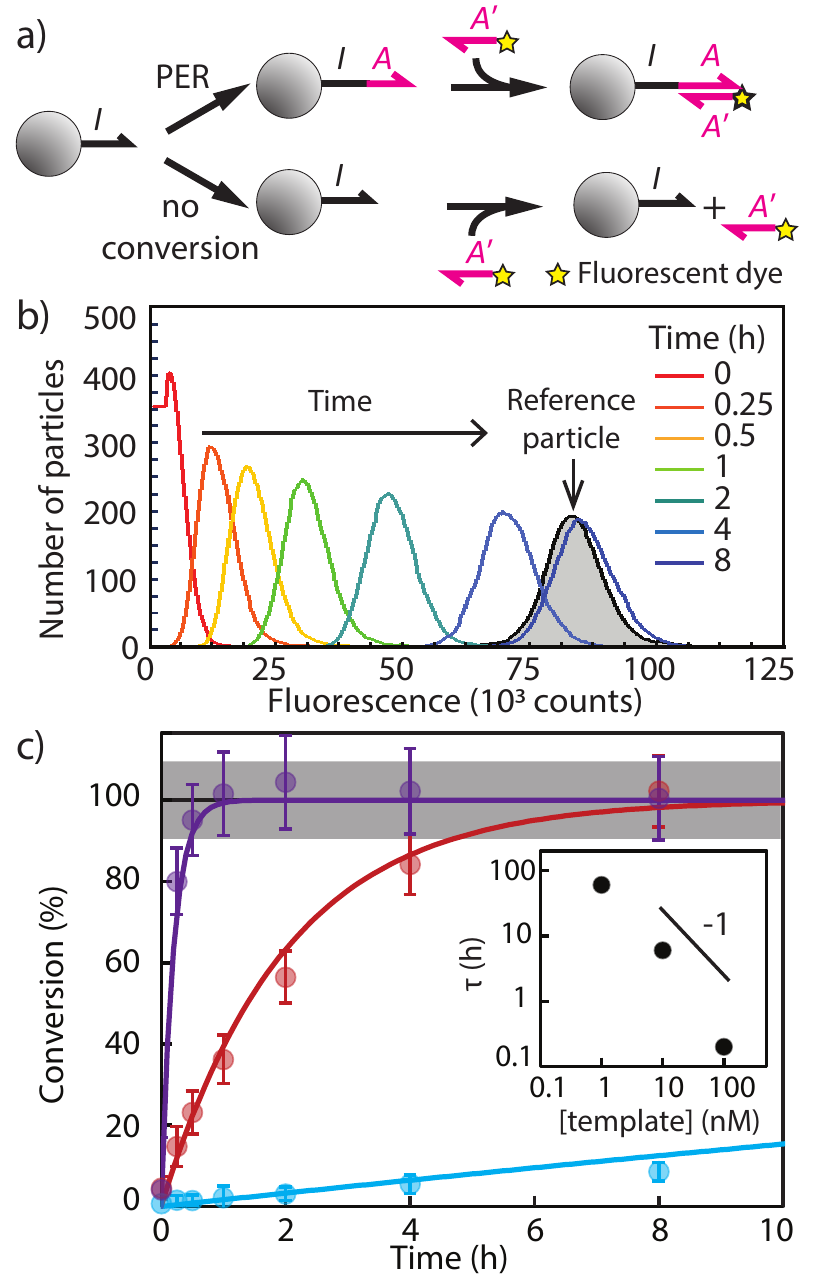}
\caption{\label{Fig1} a) Schematic of the labeling reaction used to quantify the DNA conversion on particles. The fluorescently labeled strand with sequence $A'$ is complementary to the DNA on PER-edited particles, but not to the input sequence so that the amount of fluorescence indicates the degree of conversion. b) Distribution of the single-particle fluorescence of DNA-coated particles after increasing reaction times, as measured using flow cytometry. The average fluorescence is a measure of conversion. The gray shaded region indicates the fluorescence of reference particles to which the sequence $A'$ was attached through click chemistry. Details of flow cytometry experiments are provided in the Supporting Information (SI section 1.4). Each histogram represents ten thousand particles. The template concentration was $10~\nM$.  c) PER conversion as function of time for $1~\nM$ (blue), $10~\nM$ (red), $100~\nM$ (purple) template. Higher template concentrations lead to faster conversion. Error bars represent the standard deviations of the fluorescence distributions. The particles are $600~\nm$ in diameter. The inset shows the typical reaction time, $\tau$, as a function of template concentration. The inset shows that the typical reaction time $\tau$ scales linearly with the inverse template concentration.}
\end{figure}

We first ask whether the primer exchange reaction can be used to append new sequence domains onto DNA-coated colloids. The primer exchange reaction (schematically depicted in Figure 2) involves the reversible hybridization of a single-stranded input, $I$, to the input-binding domain, $I'$, of a catalytic hairpin. When $I$ and $I'$ are bound, DNA polymerase produces the complement to the hairpin's template sequence, $A'$, appending the $A$ domain onto the input strand, resulting in a longer single-stranded output, $IA$ (Fig. 2a). The catalytic hairpin strand, which is only weakly bound to the output strand, is eventually released and can bind another input strand.

The particles whose DNA we set out to ``edit" using PER are $600$-$\nm$ and $1$-$\um$-diameter polystyrene colloids with single-stranded DNA grafted onto their surface \textit{via} the click chemistry method developed by Oh~\textit{et al.}~\cite{Oh2015} (detailed methods in SI). The grafted sequence consists of a 40-nucleotide poly-T spacer followed by a 9-nucleotide input domain, $I$. We measured a grafting density of $3.6 \pm 0.2 \times10^4$ strands/$\um^2$ on these particles (Fig. S4).

To test whether PER could append a new domain onto the DNA on the $500~\nm$ particles, we mixed the particles at $0.1\%$~(v/v) with $1-100~\nM$ hairpin strand, $100~\uM$ of each nucleotide triphosphate, and $0.13~$U$/\uL$ DNA polymerase, and let the reaction proceed at room temperature for $1$ hour. After the reaction, we washed the particles by centrifugation and resuspension. See SI for details of the synthesis and DNA sequences.

To quantify the PER conversion of DNA on the DNA-coated colloids, we added fluorescently labeled strands to the particles after the reaction and measured the fluorescent signal of ten thousand individual particles using flow cytometry (detailed methods in SI). The fluorescently
labeled DNA strands had sequence $A'$ and thus could only bind to PER-edited particles. Therefore, the fluorescent signal of each particle is a measure of the fraction of its DNA that has been converted (Fig. 3a). We determined the percent yield of the reaction by comparing the fluorescence intensity of PER-edited particles to that of reference particles to which the sequence $IA$ (the target sequence of the PER reaction) was attached directly \textit{via} click chemistry (gray shaded curve in Fig. 3b). When $10~\nM$ hairpin was used in the PER reaction, conversion of the DNA on the particles was complete after 8 hours (Fig. 3b).

Measurements of particle fluorescence after increasing reaction times indicated that the average conversion
per particle increases monotonically until complete conversion is reached. Notably, when the conversion is partial, similar fractions of DNA on each particle are converted. In other words, no two sub-populations exist of entirely unconverted and entirely converted particles (Fig. 3b). This observation suggests that by tuning the reaction time, a controllable fraction of the DNA on DNA-coated colloids can be edited with PER.

To test whether we could tune the conversion rate, we varied the concentration of catalytic hairpin used in the PER reaction. Figure 3c shows the conversion as a function of the reaction time for hairpin concentrations of $1~\nM$, $10~\nM$, and $100~\nM$. We found that the two highest concentrations reach complete conversion with a rate that increases approximately linearly with hairpin concentration. The time to complete conversion was 1 hour with $100~\nM$ hairpin and 8 hours with $10~\nM$ hairpin (Fig. 3c). With $1~\nM$ hairpin, the reaction did not go to completion within 7 days (Fig. S1). Fitting the measured conversion as a function of time to a single exponential yielded estimates of the typical reaction time where $1/e$ of the reactant was converted (inset Fig. 3c). We measured values that are a factor of 3 larger than predictions based on the rates of the PER reaction in solution (Fig. S1)~\cite{Moerman2021}. The decreased PER rate when the substrate DNA is grafted onto colloidal surfaces compared to DNA free in solution is likely due to the steric hindrance of the DNA polymerase in the dense DNA coating.



To check whether any unintended side products formed, we also performed PER on particles coated with streptavidin to which biotinylated DNA was attached (Fig. S2). The streptavidin-biotin bond can be broken by heat denaturing streptavidin at $95^o$C in $50\%$ formamide solution~\cite{Tong1992}. Using this method, we removed the DNA from the surfaces of the particles after the primer exchange reaction and analyzed the product sequences using gel electrophoresis. From this experiment we learned that only DNA strands with lengths that correspond to the lengths of the reactant and
product sequences were present on the particles after the reaction. No significant concentrations of unintended side products were formed (Fig. S2).

\subsection{Crystallization}
\begin{figure*}
\centering
\includegraphics[scale=1]{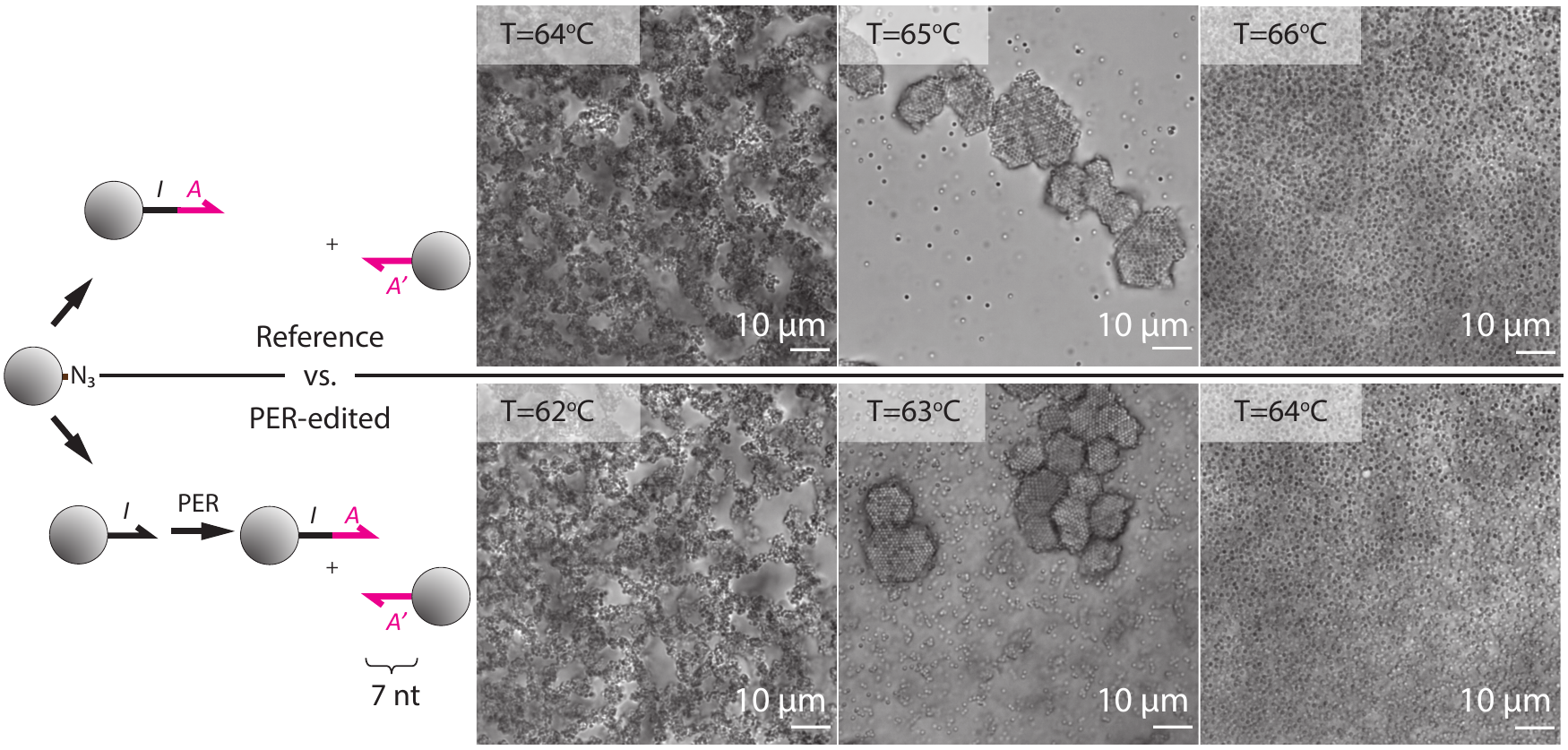}
\caption{\label{Fig1} The self-assembly of PER-edited particles (bottom row) is compared to that of reference particles (top row). When mixed with particles coated with the complementary DNA sequence, both the reference and PER-edited particles formed random aggregates below the melting temperature, crystallized when held near the melting temperature, and were unasssembled above the melting temperature. The complementary sequence, $A'$, contains 7 consecutive bases complementary with sequence $A$. All particles are $600~\nm$ in diameter. The images were recorded with an oil-immersion objective which increases thermal contact with the sample so the actual sample temperature is lower than the reported temperature.}
\end{figure*}

Our PER-editing method is intended to produce self-assembly building blocks, so we next asked whether PER-edited particles have similar assembly properties to the reference particles (to which the full sequence is
synthesized and then grafted to the particles). Typically, when a binary system of colloids coated with complementary single-stranded DNA sequences is combined, the particles form aggregates below a certain transition temperature, called the melting temperature. Above the melting temperature the colloids are unaggregated and form a stable dispersion\cite{Rogers2016}. The melting temperature increases with the hybridization free energy of the DNA strands involved in the inter-particle binding and with their grafting density\cite{Dreyfus2009}, so if the PER-edited particles and the reference particles have the same grafting density and sequence, as we expect based on the flow cytometry data, their melting temperature should also be the same. 

\begin{figure}
\centering
\includegraphics[scale=1]{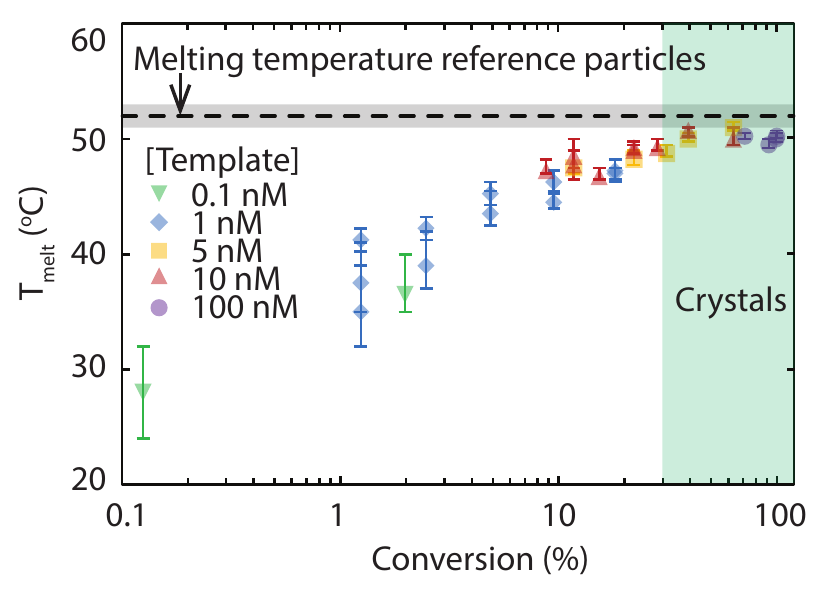}
\caption{\label{Fig1} The melting temperature as a function of approximate conversion. The approximate conversions are calculated using the fits in Fig. 3. The melting temperature increased with conversion and plateaued near the melting temperature of the reference particles (dashed black line). The green area indicates conversions for which we observed that the particles could crystallize. Below $30\%$ conversion, particles did not crystallize even if they were kept near the melting temperature. The melting temperature of unedited particles with the complementary particles is approximately $23^o$C, due to partial sequence complementarity between $I$ and $A'$ (no more than 2 sequential bases).}
\end{figure}

To determine the melting temperature we prepared samples containing either the reference particles or the PER-edited particles, and particles to which DNA containing a 7-nucleotide complementary domain was attached, which we call co-assemblers. We placed these samples on a custom-built heating element (SI section 1.3) on a microscope and found that both samples had aggregated. We then heated the sample slowly to the melting temperature, \textit{i.e.} the temperature at which approximately half the particles were part of aggregates and half were dispersed. Measurements of the fraction of single particles as function of the temperature are shown in the Supplementary Information (Fig. S6). While observing the behavior of particles during heating with a 60x oil-immersion lens, the thermistor on our our heating stage showed that aggregates of PER-edited particles and their co-assemblers melted at $63^o$C and aggregates of the reference particles and their co-assemblers at $65^o$C (Fig. 4). These measurements indicate that both particles have similar binding free energies. The 60x oil immersion objective changes the thermal load of the sample, so that the actual temperature of the sample may be lower than that reported by the temperature controller. Indeed, with a 40x air objective we found melting temperatures around $52^o$C for reference particles and $50^o$C for PER-edited particles.

Below the melting temperature, the structure that corresponds to a global minimum in the free energy landscape for a binary mixture of DNA-coated colloids is a crystal lattice, isostructural to cesium chloride\cite{Wang2015b}, that maximizes the number of contacts between the complementary particles. However, this equilibrium structure is kinetically difficult to reach, and only accessible if the particles can roll on the surface of their neighbours after binding, which requires both a high density and a homogeneous distribution of grafted DNA\cite{Moon2018, Hensley2022}. Both reference particles and input particles for PER are produced in a way that results in a DNA coating that facilitates crystallization, so we asked whether the ability to crystallize is maintained after PER.

We found that both the reference particles and the PER-edited particles crystallized readily when kept near the melting temperature (Fig. 4), indicating that the PER-edited particles are suitable building blocks for equilibrium self-assembly.

The flow cytometry data in Figure 3 show that a specific fraction of DNA on each particle can be converted by choosing an appropriate template concentration and
reaction time. To check if such partially converted particles are also suitable for self-assembly we asked how the melting temperature scales with the percentage of DNA on particles that is converted and what conversion is required for the particles to crystallize. To this end, we prepared PER-edited particles with conversion percentages ranging from 0.1\% to 100\% by varying the reaction time and template concentration. We mixed these partially converted particles with co-assemblers, similar to the experiment in Fig 4, and measured the melting temperatures. 

As the conversion increases from $0\%$, the melting temperature also increases until the melting temperature of the reference particles, $52^o$C, is reached at approximately $40\%$ conversion. Above $40\%$ conversion the melting temperature plateaued (Fig. 5). The observed increase of the melting temperature is consistent with the increase in melting temperature as a function of grafting density observed in earlier work\cite{Dreyfus2009,Geerts2010}, but the apparent plateau at $40\%$ is surprising. A notable difference with previous studies that could explain this observation is that we only varied the conversion of one of the two binding partners, and kept the grafting density of the other binding partner constant, so that the average grafting density of DNA containing sticky ends involved in the two-particle interaction goes from $50\%$ to $100\%$ as the conversion goes from $0\%$ to $100\%$. We also found that only particles with conversions over $30\%$ could crystallize with their binding partners. Below that conversion random aggregates formed even at the melting temperature consistent with the notion that crystallization requires a threshold grafting density\cite{Moon2018}.

The collapse of the measured melting temperatures as a function of fractional PER conversion for different hairpin concentrations onto a single curve shows that the fraction of DNA that has been converted is the only factor that determines the melting temperature of the PER-edited particles and how that conversion is reached does not affect the outcome. These findings show that particles with controllable conversion and---by extension---controllable melting temperature can be prepared by tuning the reaction time and catalyst concentration. 

The propensity of PER-edited particles to crystallize also depended on the PER conditions. Letting the reaction go for longer than necessary to reach $100\%$ conversion, or using more than $0.13~$U$/\uL$ Bst DNA polymerase resulted in particles that displayed non-specific aggregation even well above the melting temperature and did not crystallize (Fig. S3). The non-specific aggregation could be due to slow primer-independent or template-independent polymerization reactions~\cite{Rolando2020,Zyrina2014}.

\subsection{Reprogramming binding specificity}
\begin{figure} [t]
\includegraphics[scale=1]{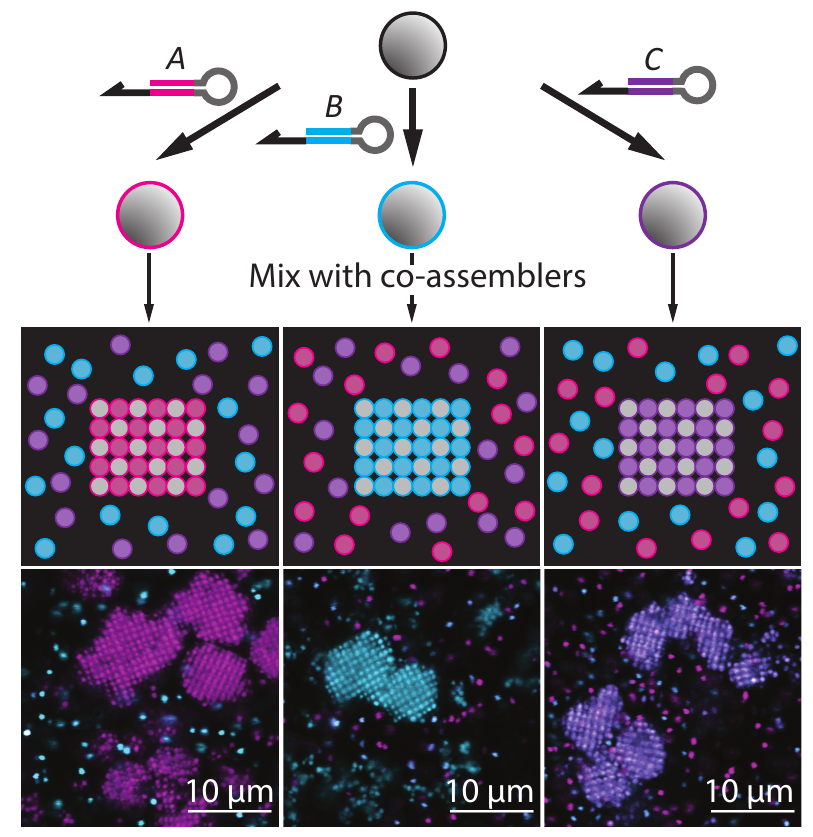}
\caption{\label{Fig4} PER converts generic DNA-coated particle into three building blocks for self-assembly with differing binding specificities. The input particles (not fluorescently labeled) are converted into three batches of DNA-coated particles with differing sequence: A, B, and C. Each batch is mixed with three types of fluorescently labeled particles: magenta particles are coated with sequence $A'$, cyan particles are coated with $B'$, and purple particles are coated with $C'$. The samples are annealed at the melting temperature and imaged at the melting temperature under a confocal microscope. Each type of converted particle aggregated only with their respective complementary particle. Scale bars are $10~\um$.}
\end{figure}

The key advance enabled by our synthesis method is that a single feed stock of DNA-coated colloids can be converted into multiple types of self-assembly building blocks with distinct binding specificities. To demonstrate this capability, we show that three different sequences can be appended onto the DNA on one type of ``input'' particle and that the resultant PER-edited particles have differing binding specificities (Fig. 6).

To this end we converted precursor particles with sequence $I$ into assembly building blocks with sequences $IA$, $IB$, and $IC$ using $100~\nM$ of three different templates. We then mixed the three PER-edited particles with a set of three co-assemblers---DNA-coated particles that contain DNA complementary to each of the PER-edited particles: $A'$, $B'$, and $C'$. The co-assembler particles were fluorescently labeled with a magenta dye, a cyan dye, or both the magenta and the cyan dyes (shown as purple), respectively. The PER-edited particles were not fluorescently labeled and are not visible in the images. We annealed a suspension of all four particle types at the melting temperature and imaged the resultant crystals using confocal microscopy. If the single feedstock of DNA-coated particles has successfully been converted into three distinct types of building blocks with differing binding specificities, each of the PER-edited building blocks should bind only to their target co-assembler and leave their off-target co-assembler particles free in solution.

Figure 6 shows that in each of the three samples, the PER-edited particles indeed co-crystallized only with their intended co-assembler; each of the crystals are either fully magenta, cyan, or purple. Notably, the co-assembler particles for the purple aggregates were also produced from the initial feed stock using the primer exchange reaction, showing that crystals form even if both types of DNA-coated particles in a binary system were produced using PER.

Finally, we tested whether multiple domains could be sequentially attached to a single particle. We tested the sequential attachment of up to three 11-nucleotide domains and found that the binding specificity of the particles successfully changed with each added DNA sequence (Fig. S5). However, particles to which more than one domain was added with PER were not able to crystallize with their complements.

\section{Conclusions}
We have introduced a method to rapidly and easily reprogram the binding specificity of DNA-coated colloids by appending new DNA domains onto the DNA grafted onto colloids. We showed that DNA-coated particles with differing binding specificities could be prepared from a single feed stock by appending new domains to the DNA on the colloids and that the particles maintained their ability to crystallize after the DNA extension procedure.

Colloid science and self-assembly have a range of open challenges that require access to building blocks with many orthogonal, tunable interactions to be addressed, such as the self-assembly of finite-sized structures of arbitrary shapes and sizes~\cite{Zeravcic2014,Jana2020,Parvez2020,Liu2016} and understanding the nucleation and growth of multi-component crystals\cite{Murugan2014,Hensley2022}. 
DNA-coated particles are a logical choice of building block because their DNA-hybridization-mediated interactions enable on the order of 100 orthogonal interactions\cite{Wu2012} with tunable strengths. However, the complexity and long duration of the synthesis---and importantly the optimization---of DNA-coated colloids has long been a barrier for their use. We think that our method will help quickly produce a wide variety of different DNA-coated colloids from a single feed-stock and optimize their designs so applications and experiments requiring many different DNA-coated colloids come within reach. Our method may also help expand the use of DNA-coated particles beyond crystallization and self-assembly to sub-fields of colloid science such as gelation and rheology\cite{Cho2020,Sherman2020,Nabizadeh2021}.

Our approach could potentially be extended beyond particles, to other systems where the grafting of DNA onto an object is expensive, time-consuming, or difficult, such as cells\cite{Todhunter2015} and antibodies\cite{Hui2015}. Our method could also be extended to change the binding specificity of single probes in DNA micro-arrays\cite{Gresham2008}.

Beyond synthesis methods, this work also opens up the possibility of altering the binding properties of particles over time within one experiment, which could be a useful tool in dissipative self-assembly. Time-dependent interactions are increasingly sought after for their ability to create dynamic, reconfigurable, and adaptive structures, but currently few chemical strategies for achieving time-dependent interaction strengths are available~\cite{Klajn2018,Ravensteijn2020}. The primer exchange reaction could also be used to initiate sequential assembly stages and freeze objects into kinetically trapped structures by converting the DNA on particles at varying rates, controlled by the hairpin concentration.

\section{Acknowledgements}
We thank Hanhvy Bui for her help with flow cytometry. P.G.M. and R.S. acknowledge funding from the ARO (grant number 90089252) and the DOE (grant number 90085886). P.G.M. acknowledges funding by the American Institute of Physics through the Robert H.G. Helleman Memorial Fellowship. H.F., T.E.V., and W.B.R. acknowledge NSF funding (grant number DMR-1710112) and funding from the Brandeis Bioinspired Soft Materials MRSEC (grant number DMR-2011846). W.B.R acknowledges support from the Smith Family Foundation.

\section{Author contributions}
P.G.M., R.S., \& W.B.R. conceived the experiments. P.G.M., T.E.V., \& H.F. performed the experiments and analysis. All authors contributed to writing of the paper.


\begin{thebibliography}{30}

\bibitem{Rogers2016}
Rogers, W.B.; Shih W.M.; Manoharan, V.N. Using DNA to program the selfassembly of colloidal nanoparticles
and microparticles. \textit{Nat. Rev. Mat.} \textbf{2015}, 1, 1-14, doi:10.1038/natrevmats.2016.8

\bibitem{Mirkin1996}
Mirkin, C.A.; Letsinger, R.L.; Mucic, R.C.;
Storhoff, J.J. A DNA-based method for rationally
assembling nanoparticles into macroscopic materials.
\textit{Nature} \textbf{1996}, 382, 607–609. doi:10.1038/382607a0

\bibitem{Alivisatos1996}
Alivisatos, A.P.; Johnson, K.P.; Peng, X.; Wilson, T.E.; Loweth, C.J.; Bruchez Jr., M.P.; Schultz, P.G. Organization of `nanocrystal molecules' using DNA. \textit{Nature} \textbf{1996}, 382, 609–611. doi:10.1038/382609a0

\bibitem{Rogers2011}
Rogers, W.B.; Crocker, J.C. Direct measurements of DNA-mediated colloidal interactions and their quantitative modeling. \textit{Proc. Natl. Acad. Sci.} \textbf{2011}, 108, 15687 —15692. doi:10.1073/pnas.1109853108

\bibitem{Liu2016}
Liu, W.; Halverson, J.; Tian, Y.; Tkachenko, V.T.; Gang. O. Self-organized architectures from assorted DNA-framed nanoparticles. \textit{Nature. Chem.} \textbf{2016}, 8, 867–873. doi:10.1038/nchem.2540

\bibitem{Park2008}
Park, S. Y.; Lytton-Jean, A.K.R.; Lee, B.; Weigand, S.; Schatz, G.C; Mirkin, C.A. DNA-programmable nanoparticle
crystallization. \textit{Nature} \textbf{2008}, 451, 553–556. doi:10.1038/nature06508

\bibitem{Auyeung2014}
Auyeung, E.; Lee, T.I.N.G; Senesi, A.J.; Schmucker, A.L.; Pals, B.C.; Olvera de la Cruz, M.; Mirkin, C.A.; DNA-mediated nanoparticle crystallization into Wulff polyhedra. \textit{Nature} \textbf{2014}. 505,
73–77. doi:10.1038/nature12739

\bibitem{Rogers2015}
Rogers, W. B.; Manoharan, V. N. Programming colloidal phase transitions with DNA strand displacement. \textit{Science} \textbf{2015}, 347, 639-642. doi:10.1126/science.1259762

\bibitem{Wang2017}
Wang, Y.; Jenkins, I. C.; McGinley, J. T.; Sinno, T.; Crocker, J. C. Colloidal crystals with diamond symmetry at optical lengthscales. \textit{Nat. Commun.} \textbf{2017}, 8, 14173. doi:10.1038/ncomms14173

\bibitem{McMullen2018}
McMullen, A.; Holmes-Cerfon, M.; Sciortino, F.; Grosberg, A.Y.; Brujic, J.; Freely Jointed Polymers Made of Droplets.
\textit{Phys. Rev. Lett.} \textbf{2018}, 121, 138002. doi:10.1103/PhysRevLett.121.138002

\bibitem{Verweij2020}
Verweij, R.; Moerman, P.G.; Ligthart, N.E.G.; Huijnen, L.P.P.; Groenewold, J.; Kegel, W.K.; Van Blaaderen, A.; Kraft, D.J. Flexibility-induced effects in the Brownian motion of colloidal trimers. \textit{Phys. Rev. Research} \textbf{2020}, 2, 033136. doi:10.1103/PhysRevResearch.2.033136

\bibitem{BenZion2017}
Ben Zion, M.Y.; He, X.; Maass, C.C.; Sha, R.; Seeman, N.C.; Chaikin, P.M. Self-assembled three-dimensional chiral colloidal architecture. \textit{Science} \textbf{2017}, 358, 633-636. doi:10.1126/science.aan5404

\bibitem{Todhunter2015}
Todhunter, M.; Jee, N.; Hughes, A.; Coyle, M.C.; Cerchiari, A.; Farlow, J.; Garbe, J.C.; LaBarge, M.A.; Desai, T.A.; Gartner, Z.J. Programmed synthesis of three-dimensional tissues. \textit{Nat. Methods} \textbf{2015}, 12, 975–981. doi:10.1038/nmeth.3553

\bibitem{Cersonsky2021}
Cersonsky, R.K.; Antonaglia, J.; Dice, B.D.; Glotzer, S.C. The diversity of three-dimensional photonic crystals. \textit{Nat. Commun.} \textbf{2021}, 12, 2543. doi:10.1038/s41467-021-22809-6

\bibitem{He2020}
He, M.; Gales, J.P.; Ducrot, É.; Gong, Z.; Yi, G-R.; Sacanna, S.; Pine, D.J.; Colloidal diamond. \textit{Nature} \textbf{2020}, 585, 524–529. doi:10.1038/s41586-020-2718-6

\bibitem{Ducrot2017}
Ducrot, E.; He, M.; Yi, G. R.; Pine, D. J. Colloidal alloys with preassembled clusters and spheres. \textit{Nat. Mater.} \textbf{2017}, 16, 652–657. doi:10.1038/nmat4869

\bibitem{Oh2019}
Oh, J. S.; Lee, S.; Glotzer, S. C.; Yi, G-R.; Pine, D. J. Colloidal fibers and rings by cooperative assembly. \textit{Nat. Commun.} \textbf{2019}, 10(1), 3936. doi:10.1038/s41467-019-11915-1

\bibitem{Zeravcic2014}
Zeravcic, Z.; Brenner, M.P. Self-replicating colloidal clusters. \textit{Proc. Natl. Acad. Sci.} \textbf{2014}, 111, 1748-1753. doi:10.1073/pnas.1313601111

\bibitem{Bausch2019}
Dehne, H.; Reitenbach, A.; Bausch, A.R. Transient self-organisation of DNA coated colloids directed by enzymatic reactions. \textit{Sci. Rep.} \textbf{2019}, 9, 7350. doi:10.1038/s41598-019-43720-7

\bibitem{Moon2018}
Moon, J.; Jo, I.S.; Ducrot, E.; Oh, J.S.; Pine, D.J, Yi, G-R. DNA-Coated Microspheres and Their Colloidal Superstructures. \textit{Macromol. Res.} \textbf{2018}, 26, 1085–1094. doi:10.1007/s13233-018-6151-8

\bibitem{Kim2006}
Kim, A.J.; Biancaniello, P.L.; Crocker, J.C. Engineering DNA-Mediated Colloidal Crystallization \textit{Langmuir} \textbf{2006}, 22, 1991-2001. doi:10.1021/la0528955

\bibitem{Vitelli2013}
Di Michele, L.; Varrato, F.; Kotar, J.; Nathan, S.H.; Foffi, G.; Eiser, E. Multistep kinetic self-assembly of DNA-coated colloids. \textit{Nat. Commun.} \textbf{2013}, 4, 2007. doi:10.1038/ncomms3007 

\bibitem{Soto2002}
Soto, C.M.; Srinivasan, A.; Ratna, B.R. Controlled Assembly of Mesoscale Structures Using DNA as Molecular Bridges. \textit{J. Am. Chem. Soc.} \textbf{2002}, 124, 8508-8509. doi:10.1021/ja017653f

\bibitem{Wang2015a}
Wang, Y.; Wang, Y.; Zheng, X.; Ducrot, E.; Lee, M.G.; Yi G-R.; Weck, M.; Pine, D.J. Synthetic strategies toward DNA-coated colloids that crystallize. \textit{J. Am. Chem. Soc.} \textbf{2015}, 137 (33), 10760-10766.

\bibitem{Oh2015}
Oh, J.S.; Wang, Y.; Pine, D.J.; Yi, G-R. High-Density PEO-b-DNA Brushes on Polymer Particles for Colloidal Superstructures. \textit{Chem. Mater.} \textbf{2015}, 27, 8337–8344. doi:10.1021/acs.chemmater.5b03683

\bibitem{Oh2020}
Oh, J.S.; He, M.; Yi, G-R.; Pine, D.J. High-Density DNA Coatings on Carboxylated Colloids by DMTMM- and Azide-Mediated Coupling Reactions. \textit{Langmuir} \textbf{2020}, 36(13), 3583-3589. doi: 10.1021/acs.langmuir.9b03386

\bibitem{Wang2015b}
Wang, Y.; Wang, Y.; Zheng, X.; Ducrot, E.; Yodh, J. S.; Weck, M.; Pine, D. J. Crystallization of DNA-coated colloids. \textit{Nat. Commun.} \textbf{2015}, 6, 7253. doi:10.1038/ncomms8253

\bibitem{Fang2020}
Fang, H.; Hagan, M.F.; Rogers, W.B. Two-step crystallization and solid–solid transitions in binary colloidal mixtures. \textit{Proc. Natl. Acad. Sci.} \textbf{2020}, 117, 27927-27933. 
doi: 10.1073/pnas.2008561117

\bibitem{Hensley2022}
Hensley, A.;Jacobs, W.M.; Rogers, W.B. Self-assembly of photonic crystals by controlling the nucleation and growth of DNA-coated colloids. \textit{Proc. Natl. Acad. Sci.} \textbf{2022}, 119, e2114050118. doi.org/10.1073/pnas.2114050118

\bibitem{IDT}
Integrated DNA Technologies (IDT) provides tailored DNA sequences with DBCO modifications ont he 3' or 5' end: https://www.idtdna.com/pages/education/decoded/
article/need-a-non-standard-modification

\bibitem{Kishi2019}
Kishi, J.Y.; Schaus, T.E.; Gopalkrishnan, N.; Xuan, F.; Yin, P. Programmable autonomous synthesis of single-stranded DNA \textit{Nature Chemistry} \textbf{2018}, 10, 155-164. doi:10.1038/nchem.2872

\bibitem{Rolando2020}
Rolando, J.C.; Jue, E.; Barlow, J.T.; Ismagilov, R.F. Real-time kinetics and high-resolution melt curves in single-molecule digital LAMP to differentiate and study specific and non-specific amplification. \textit{Nucleic Acids Res.} \textbf{2020}, 48, e42. doi:10.1093/nar/gkaa099

\bibitem{Zyrina2014}
Zyrina, N.V.; Antipova, V.N.; Zheleznaya, L.A. Ab initio synthesis by DNA polymerases. \textit{FEMS Microbiology Letters} \textbf{2014}, 351, 1–6. doi:10.1111/1574-6968.12326

\bibitem{Moerman2021}
Moerman, P.G.; Gavrilov, M.; Ha, T.J.; Schulman, R. Catalytic DNA Polymerization Can Be Expedited by Active Product Release. doi:10.26434/chemrxiv-2022-3k98v

\bibitem{Tong1992}
Tong, X.; Smith, L. M. Solid-Phase Method for the Purification of DNA Sequencing Reactions. \textit{Anal. Chem.} \textbf{1992}, 64, 2672-2677. doi:10.1021/ac00046a004

\bibitem{Dreyfus2009}
Dreyfus, R.; Leunissen, M.E.; Sha, R.; Tkachenko, R.V.; Seeman, N.C.; Pine, D.J.; Chaikin, P.M. Simple Quantitative Model for the Reversible Association of DNA Coated Colloids. \textit{Phys. Rev. Lett.} \textbf{2009}, 102, 048301. doi:10.1103/PhysRevLett.102.048301

\bibitem{Geerts2010}
Geerts, N.;Eiser, E. DNA-functionalized colloids: Physical properties and applications. \textit{Soft Matter} \textbf{6}, 4647-4660.
doi:10.1039/C001603A

\bibitem{SantaLucia1998}
SantaLucia, J. Jr. A unified view of polymer, dumbbell, and oligonucleotide DNA nearest-neighbor thermodynamics. \textit{Proc. Natl. Acac. Sci. USA} \textbf{1998}, 95, 1460-1465. doi:10.1073/pnas.95.4.1460

\bibitem{Hui2015}
Hui, J.Z.; Tamsen, S.; Song, Y.; Tsourkas, A. LASIC: Light Activated Site-Specific Conjugation of Native IgGs. Bioconjug Chem. \textit{Bioconjug. Chem.} \textbf{2015} 19, 1456-60. doi:10.1021/acs.bioconjchem.5b00275

\bibitem{Jana2020}
Jana, P.K.; Mognetti, B.M. Self-assembly of finite-sized colloidal aggregates. \textit{Soft Matter} \textbf{2020}, 16, 5915-5924.
doi:10.1039/D0SM00234H

\bibitem{Parvez2020}
Parvez, M.; Zanjani, M.B.; Synthetic Self-Limiting Structures Engineered with Defective Colloidal Clusters. \textit{Adv. Funct. Mater.} \textbf{2020}, 30, 2003317.
doi:10.1002/adfm.202003317.

\bibitem{Murugan2014}
Murugan, A.; Zeravcic, Z.; Brenner, M.P.; Leibler, S. Multifarious assembly mixtures: Systems allowing retrieval of diverse stored structures. \textit{Proc. Acad. Natl. Sci.} \textbf{2015}, 112, 54-59.
doi:10.1073/pnas.1413941112

\bibitem{Wu2012}
Wu, K.; Feng, L.; Sha, R.; Chaikin, P. Polygamous Particles. \textit{Proc. Acad. Natl. Sci.} \textbf{2012}, 106, 18731-18736. 
doi:10.1073/pnas.1207356109

\bibitem{Cho2020}
Cho, J.H.; Cerbino, R.; Bischofberger, I. Emergence of Multiscale Dynamics in Colloidal Gels. \textit{Phys. Rev. Lett.} \textbf{2020}, 124, 08800.
doi:10.1103/PhysRevLett.124.088005

\bibitem{Sherman2020}
Sherman, Z.M.; Green, A.M.; Howard, M.P.; Anslyn, E.V.; Truskett, T.M.; Milliron, D.J. Colloidal Nanocrystal Gels from Thermodynamic Principles. \textit{Acc. Chem. Res} \textbf{2021}, 54, 798–807.
doi:10.1021/acs.accounts.0c00796

\bibitem{Nabizadeh2021}
Nabizadeh, M.; Jamali, S. Life and death of colloidal bonds control the rate-dependent rheology of gels. \textit{Nat. Comm.} \textbf{2021}, 12, 4274.
doi:10.1038/s41467-021-24416-x.

\bibitem{Gresham2008}
Gresham, D.; Dunham, M.J.; Bodstein, D. Comparing whole genomes using DNA microarrays. \textit{Nat. Rev. Genet.} \textbf{2008}, 9, 291–302.
doi:10.1038/nrg2335

\bibitem{Klajn2018}
De, S.; Klajn, R. Dissipative Self-Assembly Driven by the Consumption of Chemical Fuels. \textit{Adv. Mater.} \textbf{2018}, 30, 1706750.
doi:10.1002/adma.201706750

\bibitem{Ravensteijn2020}
Van Ravensteijn, B.G.P.;Voets, I.K.; Kegel, W.K.; Eelkema, R. Out-of-Equilibrium Colloidal Assembly Driven by Chemical Reaction Networks. \textit{Langmuir} \textbf{2020}, 36, 10639–10656. doi:10.1021/acs.langmuir.0c01763


\end{thebibliography}
\end {document}


\maketitle

\section{Supplementary Methods}
\subsection{Particle Synthesis}

We synthesized DNA-grafted colloidal particles following a modified version of the method described in Ref.~\cite{Oh2015}. In brief, the method is comprised of three steps: 1) Azide groups are attached to the ends of polystyrene-poly(ethylene oxide) (PS-PEO) diblock copolymers; 2) The azide-modified PS-PEO copolymers are physically grafted to the surface of polystyrene colloids; and 3) Single-stranded DNA molecules are conjugated to the ends of the grafted PS-PEO copolymers by strain-promoted click chemistry. The specific protocol we use is described below.

We made azide-modified PS-PEO by first functionalizing PS-PEO with a methanesulfonate (Ms) group and then substituting the Ms groups with azide groups. To obtain PS-PEO-Ms, we mixed 100 mg of PS-PEO, 2 mL of dichloromethane, and 42 \ul\ of triethylamine in a glass vial, and stirred the mixture for 15 minutes on ice. Next we added 23.5 \ul\ of methanesulfonyl chloride, stirred the solution on ice for 2 hours, removed it from the ice bath and stir at room temperature for 22 hours. After the reaction, we dried the solution overnight in a vacuum dessicator and washed the dried pellet twice with a mixture of 10 mL anhydrous methanol (MeOH) and 243 \ul\ of 37\% hydrochloric acid, and then twice with a solution of 3 mL of MeOH and 40 mL of diethyl ether. In each washing step, we dissolved the pellet and then precipitated the PS-PEO by placing the sample in the freezer for one hour. Then we centrifuged the solution at 4500 rpm at 2 \degree{C} for 10 minutes to form a pellet and poured off the supernatant. After washing, we dried the pellet again.

Next, we substituted the Ms groups with azide groups. We mixed 10 mg of sodium azide, 2 mL of dimethylformamide, and the dried PS-PEO pellet. We placed the solution in a 65 \degree{C} oil bath and stirred at 1500 rpm for 24 hours. After the reaction, we washed the mixture with 40 mL of diethyl ether and then with a solution containing 3 mL of MeOH and 40 mL of diethyl ether three times. We used the same washing procedure as we previously described. Then we dried the pellet overnight in a vacuum dessicator and resuspended the dried PS-PEO-N3 pellet in deionized (DI) water to a concentration of $100~\mM$.

We attached the azide-modified PS-PEO copolymer to the surface of polystyrene colloids using a physical grafting method. We first adsorbed PS-PEO-N$_3$ to the surface of polystyrene colloids by mixing 160 \ul\ of 100 mM PS-PEO-N3, 160 \ul\ of tetrahydrofuran, 40 \ul\ of deionized (DI) water, and 40 \ul\ of 10\%(v/v) 600-nm-diameter PS particles (Molecular Probes), and then vortexed the mixture for 30 minutes. Next, we diluted the mixture 10x with DI water, washed the particles with DI water five times, and concentrated the particles back to 1\%(v/v) after washing.

We dyed particles with different fluorophores so that we could distinguish between different particle species. Specifically, we labeled three types of particles, one with nile red, one with pyrromethene green, and one with both. We first made saturated solutions of fluorescent dye dissolved in toluene. We mixed 4 \ul\ of 10\% saturated nile red or 50\% saturated pyrromethene green in toluene and 400 \ul\ 1\% azide-functionalized PS particles and rotated end-over-end for 7 hours. Next we opened the sample to air and heated it in an oven at 90 \degree{C} for 12 minutes. After that, we washed the dyed particles five times in DI water by centrifugation and resuspension.

Finally, we attached DNA molecules to the azide-modified PS-PEO copolymers using strain-promoted click chemistry. We mixed $40~\uL$ of 1\% azide-functioned PS particles, $10~\uL$ of $100~\uM$ ssDNA, and $150~\uL$ 1x TE/$1~\M$ NaCl buffer containing $0.05$~wt\% Pluronic F127, and heated the sample in an oven at 70 \degree{C} for 24 hours. After the reaction, we washed the DNA-coated particles with DI water five times by centrifugation and resuspension.

\subsection{PER Reaction}
To append new sequences onto the DNA grafted to the particles, we used an adapted version of the Primer Exchange Reaction (PER) described in Ref.\cite{Kishi2019}:
First we prepared a PER reaction mixture containing final concentrations of 1x Thermopol DNA polymerase buffer (New England Biolabs, provided at 10x concentration with the DNA polymerase), $12.5~\mM$ MgCl$_2$ (New England Biolabs, provided as a $100~\mM$ solution with the DNA polymerase), and $100~\uM$ of each dATP (deoxyribose adenine triphosphate, ThermoFisher, $10~\mM$), dCTP, and dTTP. No dGTP was added to the mixture because only ACT sequences were appended onto particles and a G-C pair was used as a stop sequence (see section: ``Design Considerations"). Note, for the purple co-assembler particles in Figure 4 a different reaction mixture was prepared containing dGTP, but no dCTP. There an AGT sequence was appended onto the particle and a C-G pair was used as stop sequence.
All solutions were prepared in Millipore purified water. DNA was stored at $-20$ \degree{C} and DNA-coated particles were stored at $4$ \degree{C}

For the primer exchange reaction, we mixed $5~\uL$ of $1$~wt\% DNA-coated particle suspension to a final concentration of $0.1$~wt\%; $5~\uL$ of $1~\uM$ DNA hairpin solution to a final concentration of $100~\nM$; $25~\uL$ of the premixed PER reaction mixture; and $1~\uL$ of $8~$U/$\uL$ Bst Large Fragment DNA polymerase (New England Biolabs) to a final concentration of $0.13~$U/$\uL$; and $14~\uL$ water to a total reaction volume of $50~\uL$ in a $1~\mL$ Eppendorf tube. For Figure S1 larger concentrations of Bst DNA polymerase were used as indicated in the figures. For Figure 2 and S2 varying hairpin concentrations were used as indicated in the figures

The reaction mixture was rotated at room temperature (approximately 24 \degree{C}) for 1 hour, after which the reaction was terminated by washing. 
We found that the rotation to prevent sedimentation is not essential for reaction times under $2$ hours.  
We washed by centrifuging the particles 4 times at $3000 \times g$ for 3 minutes, removing $45~\uL$ reaction mixture each time and resuspending the particles in $50~\uL$ by adding $45~\uL$ of water. Reaction times were varied for Figures 2 and 3c as indicated in the figures.

Except for the DBCO-modified DNA, all DNA was purchased unpurified (\textit{i.e.} standard desalting) from Integrated DNA Technologies, in the ``lab ready" formulation (dissolved in IDTE buffer at $100~\uM$). The DBCO-modified DNA (Integrated DNA Technologies) was HPLC purified.

\subsection{Crystallization Experiments}

To prepare samples for crystallization we mixed 1 \ul\ of 1\%(v/v) each of two complementary DNA-coated particle types (2 \ul\ total) with 2 \ul\ of 1xTE Buffer with 500 mM NaCl, for a total sample volume of 4 \ul. We take 1.6 \ul\ of the solution and pipette it onto a plasma-cleaned 24 mm x 60 mm coverslip in the center of a thin, open ring of high vacuum grease (Dow Corning). A second piece of 15 mm x 15 mm plasma-cleaned coverslip is placed on top, making a seal with the vacuum grease. When sealing the chamber we removed as much air as possible without allowing the solution to move past the vacuum grease ring before it closes off. For longer timescale experiments, an additional seal of UV glue was used on the edges of the coverslip to prevent shear on the chamber as well as reduce further the chance for evaporation to break the grease seal. 

To control the temperature of the sample, we taped it to home-built temperature stage consisting of a peltier element with a thermistor controlled by a PID controller. We started experiments by raising the temperature to the point where all aggregates melted. The sample was left at this temperature for approximately 30 minutes until the particle density became uniform across the chamber. The temperature was then lowered in 0.5~\degree{C} steps and held for 5 minutes at each point. By looking at the fraction of particles that have aggregated at each temperature we found the melting temperature, where 50\% of particles are aggregated. To form crystals, we melted the sample and then held the temperature $\sim$0.3 \degree{C} above the melting temperature. It took about three hours for crystal domains to form.

We imaged the samples using either an inverted, brightfield microscope (Nikon Eclipse TE2000-E) or a confocal microscope (Leica SP8) equipped with 20x air, 40x air, and a 60x oil immersion objectives.

\subsection{Flow Cytometry}
Flow cytometry experiments were performed on a BD FACSCanto high-throughput analytical flow cytometer. Samples were prepared in a $1~\mL$ Eppendorf tube by diluting $1~\uL$ of $0.1$~wt\% into $100~\uL$ buffer containing $10~\mM$ Tris-HCl (pH=7.5), $1~\M$ NaCl, and $1~\mM$ EDTA and varying concentrations of fluorescently labeled DNA complementary to the DNA on the particle. In Figure 2 we added $10~\uL$ of $1~\uM$ reporter strand to a final concentration of $100~\nM$. In Figure S4, the concentrations varied as indicated in the figure.

The samples were left to equilibrate for at least one hour, then they were diluted in $400~\uL$ of the same buffer (without additional reporter strand), vortexed, and transferred to a flow cytometry tube immediately. The fluorescent signal of 10 thousand particles was collected over approximately 5 minutes, using the low flow rate setting.

Of all measured events, only the subpopulation involving single particles was selected by applying a gate based on forward and side scatter intensity. Events involving dimers, or larger aggregates were discarded. Data were analyzed in FCS Express 6 Flow Research Edition.

\subsection{Gel electrophoresis}
We measured whether PER on particles led to any unintended side products, such as shorter or longer strands than the intended product. To visualize the length of a DNA strand in an electrophoresis gel, it needs to be removed from the particle. Therefore, we PER-edited particles coated with DNA via the streptavidin-biotin bond. This bond can be broken by heat denaturing in formamide, to release the DNA after the reaction.

We first prepared DNA-coated particles from streptavidin-coated particles ($1~\um$ in diameter, Bangs Labs) by mixing $0.1$~wt\% particles with $10~\uM$ biotin-functionalized DNA in a total volume of $50~\uL$. The particles were placed on a rotator for 1 hour and washed by centrifugation and resuspension (3x, 3 min at 3000 rcf). Then the DNA on their surface was converted using PER according to the description above. The reaction was stopped and the DNA removed from the particle simultaneously by adding loading buffer in formamide and heat denaturing the sample at $95~^o$C for 5 minutes. This step deactivated the DNA polymerase and broke the biotin-streptavidin bond. The samples were then centrifuged and the supernatants were loaded into a $15~\%$ polyacrylamide gel in a bath of 1x Tris/Borate/EDTA (TBE) buffer heated to $65~^o$C. Then $100~$V was applied for 2 hours. The gels were stained with Sybr gold dye and imaged using a SynGene Genebox gel imager operated with the Genesys software. 

\break

\subsection{DNA Sequences}

Here we list the DNA sequences (Integrated DNA Technologies) that we used for this work. The sequences are color coded to match Figure 1: the input domain is in black and bold, output A and its sequence complement in red, output B and its sequence complement in blue, and output C and its sequence complement in purple.

\begin{figure}[b]
\centering
\includegraphics[scale=1.1]{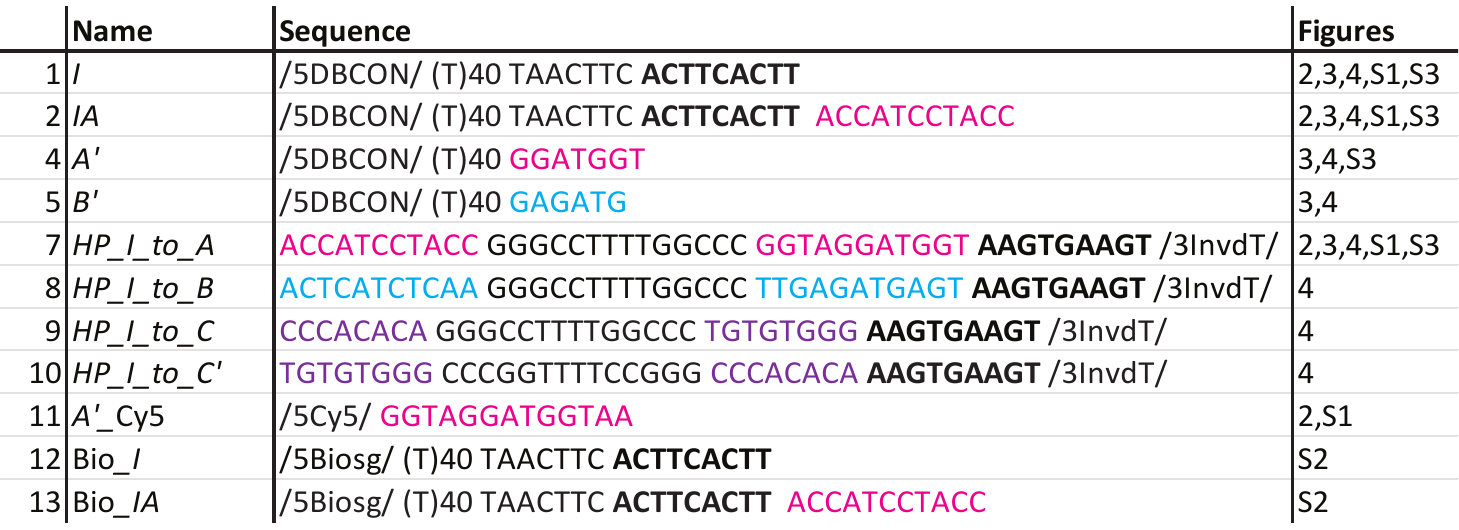}
\end{figure}

\break

\section{Design considerations}
While PER has clear advantages in the preparation of DNA-coated colloids, there are also some specific limitations, as well as considerations for designing the templates that should be taken into account. 

\subsection{Inherent limitations}
First, PER can only extend sequences from the 3' end of DNA. Whenever a 5' sticky end on the DNA is required, our method cannot be used and the particles must be synthesized using click chemistry. Second, PER can not be used to produce grafted DNA molecules with non-natural bases or chemical modifications. Third, our method only works well if the DNA on the feed stock particles already have a single-stranded domain of at least 8 bases.

\subsection{Template design principles}
\indent To design a new template, it is important to follow three design principles that relate to the three sections of the template: the single-stranded binding domain, the sequence template domain, and the stop sequence. 
\\
\indent The single-stranded binding domain is complementary to the input sequence and is responsible for hybridization to the input strand. The length of this binding domain determines the rate of the reaction. In our earlier work on PER~\cite{Moerman2021}, we found that the reaction halftime is given by $\tau= (\frac{1}{k_2}+\frac{K}{k_f}) (\frac{R_0}{C_0}+\frac{1}{K C_0})$, where the polymerization rate $k_2=3 \times 10^{-3}~\ps$, the DNA hybridization rate $k_f=3 \times 10^6 \M^{-1}\ps$, the reactant concentration $R_0~\approx 200~\nM$, the catalyst concentration $C_0$ varies, and $K$ is the equilibrium constant for catalyst-reactant binding. $K$ depends on the sequence and length of the binding domain and can be predicted using the well-established parameters of DNA hybridization thermodynamics~\cite{SantaLucia1998}. The rate is optimal for $K=\left ( \frac{k_f}{R_0 k_2}\right )^{0.5}$. At room temperature and for sequences with $0.3-0.5$ GC-content, this corresponds to an optimal domain length of 9 nucleotides.
\\
\indent The second design rule considers the template domain, which is the double-stranded domain that contains the sequence of the DNA that will be appended onto the input sequence. Long template strands likely result in slow reaction kinetics, so the addition of domains longer than ten nucleotides should be done in consecutive PER reaction steps (Supp. Fig. 5). Kishi~\textit{et al.} showed that multiple PER conversions can be done in a one-pot reaction using multiple hairpins~\cite{Kishi2019}. We anticipate that this same scheme could be used to make DNA-coated colloids with appended domains that are longer than ten nucleotides, but we have not tested it. We also expect that appending shorter-than-ten-nucleotide domains is not a problem, but have not tested it.
\\
\indent The final design rule considers the DNA stop sequence. The DNA template used for PER requires a stop sequence that the DNA polymerase cannot copy. One useful trick is to append a DNA sequence that contain only 3 out of the 4 nucleotides, which allows you to use a stop sequence that is the fourth nucleotide. For example, if one adds a sequence only containing A's, C's, and T's, the stop sequence can be a G-C pair with the G on the non-template strand. In that case, the reaction mixture should not contain any dGTP so that the DNA polymerase stops copying the sequence at the G-C pair. We used this method for the experiments presented in this paper. It is also possible to append sequences containing all four nucleotides by using a non-natural base-pair, such as iso-dC and iso-dG or methylated RNA bases as a stop sequence. However, strands containing these non-natural bases are more expensive and time-consuming to produce.

\break

\begin{figure}[t]
\centering
\includegraphics[scale=1]{SI_fig1_model&longtimeconversion2.pdf}
\caption{a) Conversion of particle-grafted DNA as a function of time due to the Primer Exchange Reaction. The samples contain $100~\nM$ hairpin (purple), $10~\nM$ hairpin (red), and $1~\nM$ hairpin (blue). Where higher catalyst concentrations lead to complete conversion in $1$ and $8$ hours respectively, the sample with $1~\nM$ hairpin does not reach complete conversion within one week. b) The same data as in (a) plotted as function of a normalized time (time multiplied by hairpin concentration). The higher concentration data collapse onto the same curve, but the low concentration data result in lower than expected conversion. This is probably due to a small amount of template sticking to the reaction tube. If $0.5~\nM$ is removed from the reaction because it sticks to the tube, that hardly affects rates at  hairpin concentrations of $10~\nM$ and $100~\nM$, but it has a large effect on reaction rates when only $1~\nM$ hairpin is present. The black line is a fit of $P/P_0=1-\exp{(\tau/t)}$. Based on our earlier work~\cite{Moerman2021}, we predicted that under the experimental conditions ($R_0=36~\nM$, $K=0.5~\nM^{-1}$, $k_2=2\times 10^{-3} \ps$, $k_f=3 \times 10^6 \M^{-1}\ps$) $\tau=446$ min $\nM$. The fit yields $\tau=1.2\times 10^3$ min $\nM$, suggesting PER proceeds slightly slower on the surface of colloids than in bulk.}
\end{figure}

\begin{figure} [t]
\centering
\includegraphics[scale=1]{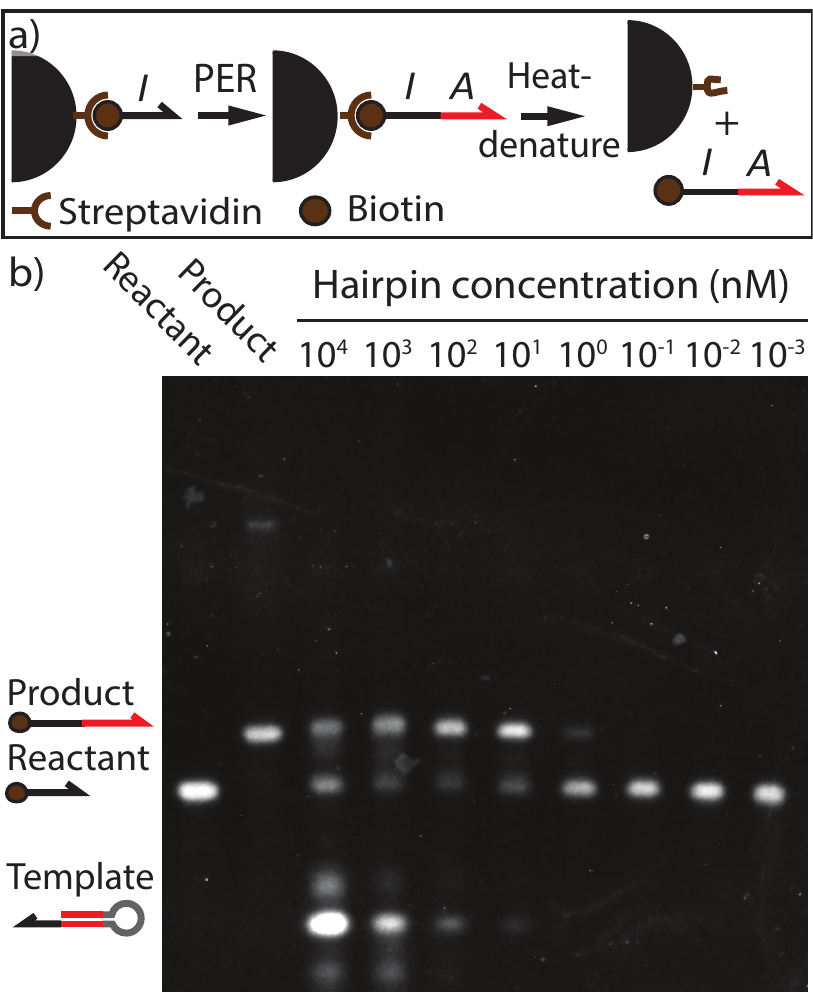}
\caption{Electrophoresis gel of DNA grafted onto particles before and after PER. Samples were prepared by attaching Biotinylated DNA to Streptavidin-coated particles, then performing the Primer Exchange Reaction, and finally removing the DNA from the particles under denaturing conditions. The first two lanes contain samples in which no PER was performed, but reactant and product strands were attached to the particle. The lanes indicate that the strands can be removed from the particles and imaged in an electrophoresis gel. The other lanes show the DNA that results after $2$ hours of PER with decreasing template concentrations. Samples with higher template concentrations result in more conversion, but no complete conversion is observed for these particles at any conditions. Notably, in each sample there are only two bands observable: the reactant and the product. No unintended side products are formed at any conditions. In the samples with higher template concentrations, the template band is also visible.}
\end{figure}

\begin{figure} [t]
\centering
\includegraphics[scale=1]{SI_fig3_melting&crystallization_2.pdf}
\caption{Melting curves of DNA-coated particles with their binding particles that have been prepared under various PER conditions. The black line represents reference particles to which the assembly DNA is attached directly via click chemistry. This sample showed the characteristic steep melting curve where all particles are singlets only a few degrees above the melting temperature (iii). The red line represents a sample prepared using PER where we used the DNA polymerase concentration that was originally used in the paper by Kishi \textit{et al.}~\cite{Kishi2019}. We observed non-specific aggregation that persisted well above the melting temperature (iv). Washing the particles repeatedly in water to remove excess polymerase reduced the problem, but small clusters continued to persist and the sample did not crystallize (ii). Preparing samples using a 5-fold reduced DNA polymerase concentration (blue line) resulted in a sharp melting curve again and enabled crystallization (i).}
\end{figure}

\begin{figure} [t]
\centering
\includegraphics[scale=1]{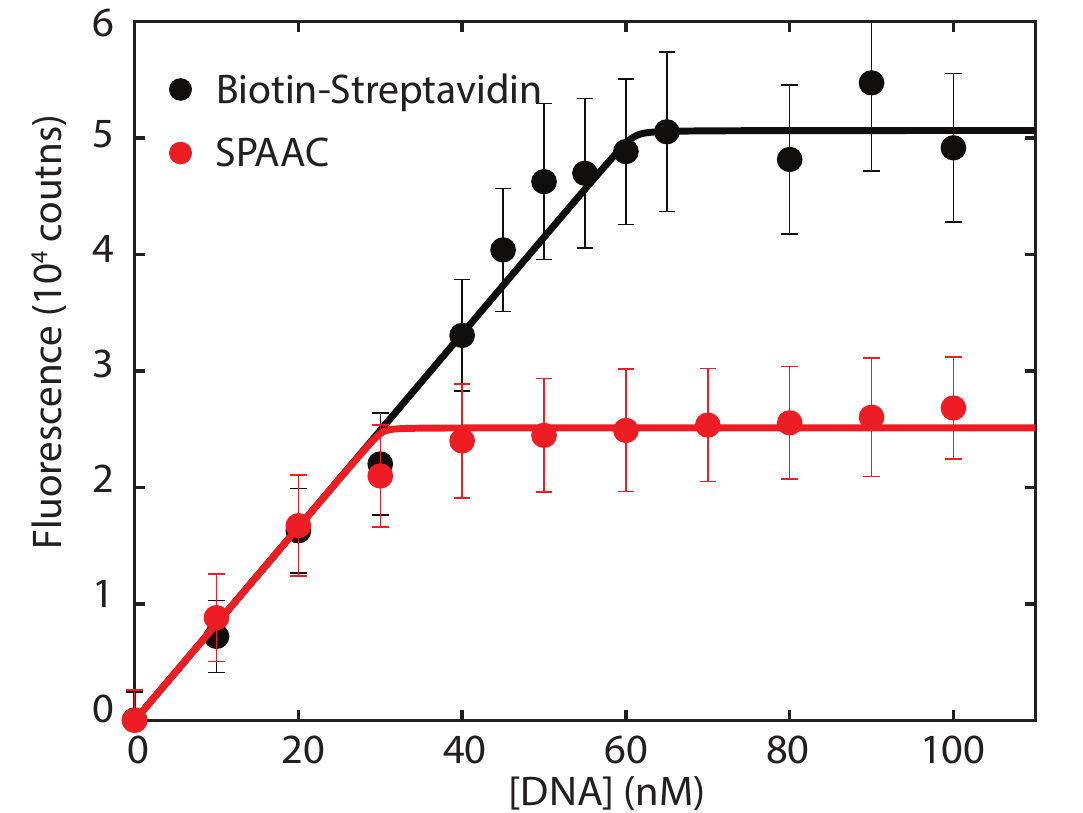}
\caption{The binding capacity of fluorescently labeled strands onto a DNA-coated particle is measured as a proxy for the DNA grafting density. We measured the fluorescent signal of DNA-coated particles mixed with increasing amounts of Hex-labeled complementary DNA. The fluorescent signal linearly increased until it plateaus. The transition to a plateau indicates saturation of the DNA grafted onto the colloids. We fit the data with equation $F= \frac{q}{2 K c_p} \left (c_f K + c_p N K + 1 \right) - \sqrt{\left( c_f K + c_p N K + 1 \right)^2 - 4 K^2 c_f c_p N}$ that follows from equilibrium binding. Here $F$ is the fluorescent signal in count, $q=0.2$ is a fit parameter that represents the fluorescent signal per fluorescently labeled DNA strand $c_p=2.6 \pm 0.2 \times 10^{-4} \nM$ is the particle concentration, $c_f$ is the fluorescently-labeled DNA concentration in nanomolar, $K=139.5~\nM^{-1}$ is the equilibrium constant for the hybridization reaction between a fluorescently labeled DNA strand and a DNA strand on the particle, and $N$ is the number of DNA strands per particle. After fitting we find that $N=2.3 \pm 0.2 \times 10^5$ strands per particle on the DNA-coated particles prepared using Biotion-Streptavidin chemistry and $N=1.14 \pm 0.05 \times 10^5$ strands per particle on the DNA coated particles prepared using click chemistry.}
\end{figure}

\begin{figure} [t]
\centering
\includegraphics[scale=1]{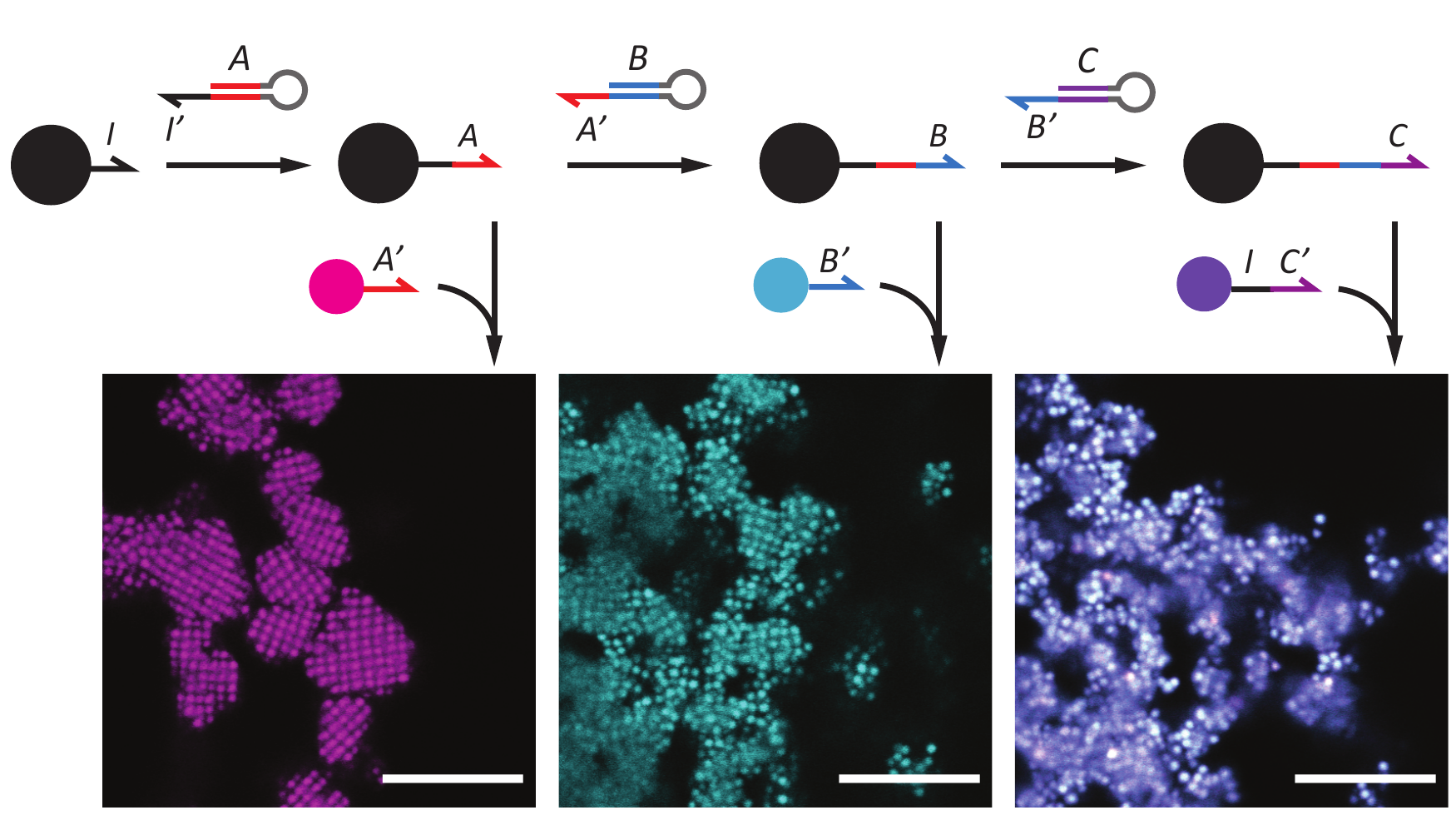}
\caption{Multi-domain DNA strands can be grown onto colloids via sequential PER reactions. Here the generic input sequence, $I$, is first extended to form output sequence $IA$. Then the $IA$ sequence is extended with a B domain. Finally the $IAB$ sequence is extended with a $C$ domain. Particles with the $IA$ sequence aggregate with $A'$ co-assemblers, particles with the $IAB$ sequence aggregate with $B'$ co-assemblers, and particles with the $IABC$ sequence aggregate with $C'$ co-assemblers, indicating that the conversions were successful.}
\end{figure}

\clearpage

\begin{figure} [t]
\centering
\includegraphics[scale=1]{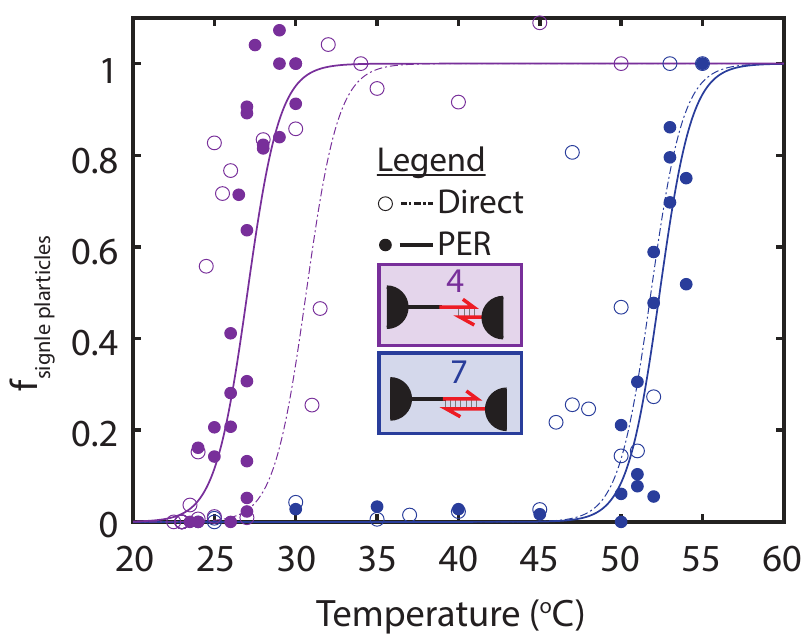}
\caption{The fraction of non-aggregated particles as a function of temperature for both the reference particles (hollow circle, dashed line) and particles converted by PER (filled circle, continuous line). The lines are fits to the experimental data. The width of the melting curve is on the order of two degrees, both for the PER particles and the reference particles. The melting curves match well in the system in which the co-assembling particles have 7 complementary bases (blue) and are within the error in the system in which they have 4 complementary bases (purple).}
\end{figure}